\documentstyle[12pt]{article}
\newlength{\extraspace}
\newlength{\extraspaces}
\newcommand{\be}{\begin{equation}
\addtolength{\abovedisplayskip}{\extraspaces}
\addtolength{\belowdisplayskip}{\extraspaces}
\addtolength{\abovedisplayshortskip}{\extraspace}
\addtolength{\belowdisplayshortskip}{\extraspace}}
\newcommand{\ee}{\end{equation}}
\newcommand{\ba}{\begin{eqnarray}
\addtolength{\abovedisplayskip}{\extraspaces}
\addtolength{\belowdisplayskip}{\extraspaces}
\addtolength{\abovedisplayshortskip}{\extraspace}
\addtolength{\belowdisplayshortskip}{\extraspace}}
\newcommand{\ea}{\end{eqnarray}}
\begin{document}
\setlength{\baselineskip}{6mm}
\vspace{7mm}
\begin{center}
{\large \bf  New Einstein-Hilbert type action of  space-time and matter  \\
-Nonlinear-supersymmetric general relativity theory-
} \\[15mm]
{\sc Kazunari Shima}
\footnote{
\tt 
e-mail: shima@sit.ac.jp \\
Based on the talk at European Physical Society Conference on High Energy Physics (EPS-HEP2019),10-17 July, 2019, Ghent, Belgium}
\\[5mm]
{\it Laboratory of Physics, 
Saitama Institute of Technology \\
Fukaya, Saitama 369-0293, Japan} \\[20mm]
\begin{abstract}
The geometric argument of the general relativity principle can be carried out on (unstable) Riemann space-time just inspired by nonlinear representation of supersymmetry(NLSUSY), where tangent space is specified by Grassmann degrees of freedom $\psi$ for SL(2,C) besides the ordinary Minkowski one $x^{a}$ for SO(1,3)  and gives straightforwardly new Einstein-Hilbert(EH)-type action with global NLSUSY invariance(NLSUSYGR))  equipped with the cosmological term. Due to the NLSUSY nature of space-time NLSUSYGR would collapse(Big Collapse) spontaneously to ordinary E-H action of graviton, NLSUSY action of Nambu-Goldstone fermion $\psi$ and their gravitational interaction. Simultaneously the universal attractive gravitational force would constitute the NG fermion-composites corresponding to the eigenstates of liner-SUSY(LSUSY)  super-Poincar\'{e}(sP)  symmetry of space-time, which gives a new paradigm for the unification of space-time and matter. 
By linearizing NLSUSY we show that the standard model(SM) of the low energy particle physics can emerge in the true vacuum of NLSUSYGR as the NG fermion-composite massless eigenstates of LSUSY sP algebra of space-time symmetry, which can be understood as the ignition of the Big Bang and continues naturally to the standard Big Bang model of the universe. 
NLSUSYGR can bridge naturally the cosmology and the low energy particle physics and provides new insights into unsolved problems of cosmology, SM and  mysterious relations between them, e.g. the space-time dimension {\it four}, the origin of SUSY breaking, the dark energy and dark matter,  the dark energy density$\sim$( neutrino mass$)^{4}$, the tiny neutrino mass, the three-generations structure of quarks and leptons, the rapid expansion of space-time,  the magnitude of bare gauge coupling constant,  etc..   
\end{abstract}
\end{center}

\newpage
\noindent
\underline{\bf  1. Three-generations strucure} 　\\[3mm]
\indent 
 Supersymmetry (SUSY)\cite{WZ1,VA,GL}
related naturally to space-time symmetry is promissing for the unification of general relativity and the low enegy SM in {\it one} single irreducible representation of the symmetry group\cite{MG}. 
We have found by group theoretical arguments that 
among all $SO(N)$ super-Poincar\'e (sP) groups  the $SO(10)$ sP group decomposed as 
${N = \underline{10} = \underline{5}+\underline{5^{*}}}$ under $SO(10) \supset SU(5)$
may be a unique and minimal group which accomodates all observed particles including graviton in {\it a single} irreducible representation of  $N$ {\it linear(L)} SUSY($N>10$ is excluded phenomenologically.)\cite{KS1, KS3}. 
($N>10$ is excluded phenomenologically.)
In this case  
10 supercharges $Q^{I}, (I=1,2, \cdots. 10)$ are embedded as follows:  %
{{} {{} $\underbar{10}_{SO(10)}=\underbar{5}_{SU(5)}+{\underbar{5}^{*}}_{SU(5)}$}}, \   
{{} {${} {\underbar5}_{SU(5)} =[ \  \underbar3^{*c}, {\underbar1^{ew}}, ({e \over 3},{e \over 3},{e \over 3}):Q_a(a=1,2,3)  \ ] + [\  \underbar1^{c}, {\underbar2}^{ew}, (-e, 0): Q_m(m=4,5) \ ]$}}, .\ 
i.e., 
${  \underbar5_{SU(5)GUT}}$ represents  { [$Q_{a}$: $\bar d$-type,{}
$Q_{m}$: lepton-type]}  supercharges.{}   
The massless state $|{}{h}>$  of {{} gravity supermultiplet}  with $n$-supercharge of SO(10) sP with CPT conjugate are specified by the helicity $h=({{} 2}-{n \over 2})$ 
and the dimension
${\underline d}_{[n]}=\underline{{{10!} \over {n!(10-n)!}}}(n=0,1, ,10)$ as tabulated below\cite{KS3}.    
\begin{table}[htb]
\begin{center}
\begin{tabular}{|c||c|c|c|c|c|c|c|} \hline
\hspace{0mm} $|h|$ \hspace{0mm} & \hspace{0mm} $3$ \hspace{0mm} & 
\hspace{0mm} ${5 \over 2}$ \hspace{0mm} & 
\hspace{0mm} ${} 2$ \hspace{0mm} & \hspace{0mm} ${3 \over 2}$ \hspace{0mm} & 
\hspace{0mm} ${1}$ \hspace{0mm} & \hspace{0mm} ${1 \over 2}$ \hspace{0mm} & 
\hspace{0mm} ${0}$ \hspace{0mm}   \\ \hline 
\raise2ex\hbox{$
\begin{array}{c}
{{} {} }    \\[2mm]
{\underline {d}_{[n]} }  \\
\end{array} 
$}  
& 
\raise2ex\hbox{$
\begin{array}{c}
{{} {} }    \\[2mm]
{\underline {1}_{[10]} }  \\
\end{array} 
$}  
& 
\raise2ex\hbox{$
\begin{array}{c}
{} \\[2mm]
{\underline {10}_{[9]}} \\
\end{array} 
$} 
&
\raise2ex\hbox{$
\begin{array}{c}
{\underline {{} 1}_{[0]} }    \\[2mm]
{\underline {45}_{[8]} }  \\
\end{array} 
$} 
& \raise2ex\hbox{$
\begin{array}{c}
{\underline {10}_{[1]}}    \\[2mm]
{\underline {120}_{[7]} }  \\
\end{array} 
$} 
& \raise2ex\hbox{$
\begin{array}{c}
{\underline {45}_{[2]}}    \\[2mm]
{\underline {210}_{[6]}}  \\
\end{array} 
$} 
& \raise2ex\hbox{$
\begin{array}{c}
{\underline {120}_{[3]}}    \\[2mm]
{\underline {252}_{[5]}}  \\
\end{array} 
$} 
& \raise2ex\hbox{$
\begin{array}{c}
{\underline {210}_{[4]}}    \\[2mm]
{\underline {210}_{[4]}}  \\
\end{array} 
$} 
\\ \hline
\end{tabular}
\end{center}
\end{table}
\noindent

To see low energy massive states we assume $SU(3) \times SU(2) \times U(1)$ invariant   {\it maximal}   superHiggs-like mechanism among helicity states, 
i.e., all redundant high helicity states for SM become massive by absorbing  lower helicity states (and  decoupled)  in {\it SM invariant way}. 
%
Many lower helicity states with $h=0, {1 \over 2}$ disappear from the physical states and the surviving states reveal a interesting structure: 
{Spin ${1 \over 2}$ state survivours} are shown in the table 
(tentatively  as Dirac particles).  
%
\begin{table}
\begin{center}
\begin{tabular}{|c|c|c|} \hline
\hspace{0mm} $SU(3)$ \hspace{0mm} & \hspace{0mm} $Q_e$ \hspace{0mm} & \hspace{0mm} 
$SU(2) \otimes U(1)$ \hspace{0mm} \\ \hline 
\underline{{} 1} & 
$
\begin{array}{c}
0 \\[2mm]
-1 \\[2mm]
-2
\end{array} 
$ 
& \raise2ex\hbox{$\left(
\begin{array}{c}
{ \nu_e }    \\[2mm]
{{} e }  \\
\end{array} 
\right)$} 
\raise2ex\hbox{$\left(
\begin{array}{c}
{{} \nu_\mu} \\[2mm]
{{} \mu }
\end{array} 
\right)$} 
\raise2ex\hbox{$\left(
\begin{array}{c}
{{} \nu_\tau}   \\[2mm]
{{} \tau} 
\end{array} 
\right)$} 
\lower4ex\hbox{$
\begin{array}{c}
\left(
E
\right)
\end{array} 
$}
\\ \hline
\underline{{} 3} & 
$
\begin{array}{c}
{5/3} \\[2mm]
{2/3} \\[2mm]
-{1/3} \\[2mm]
-{4/3}
\end{array} 
$ 
& \lower1ex\hbox{$\left(
\begin{array}{c}
{{} u }   \\[2mm]
{{} d }   \\
\end{array} 
\right)$} 
\lower1ex\hbox{$\left(
\begin{array}{c}
{{} c}  \\[2mm]
{{} s }  
\end{array} 
\right)$} 
\lower1ex\hbox{$\left(
\begin{array}{c}
{{} t } \\[2mm]
{{} b }   
\end{array} 
\right)$} 
\lower5ex\hbox{$\left(
\begin{array}{c}
h \\[2mm]
o
\end{array} 
\right)$}
\raise4ex\hbox{$\left(
\begin{array}{c}
a \\[2mm]
f
\end{array} 
\right)$}
\raise4ex\hbox{$\left(
\begin{array}{c}
g \\[2mm]
m
\end{array} 
\right)$}
\lower3ex\hbox{$\left(
\begin{array}{c}
r \\[2mm]
i \\[2mm]
n
\end{array} 
\right)$}
\\ \hline
\underline{{} 6} & 
$
\begin{array}{c}
{4/3} \\[2mm]
{1/3} \\[2mm]
-{2/3}
\end{array} 
$ 
& $\left(
\begin{array}{c}
P \\[2mm]
Q \\[2mm]
R
\end{array} 
\right)$
$\left(
\begin{array}{c}
X \\[2mm]
Y \\[2mm]
Z
\end{array} 
\right)$
\\ \hline
\underline{{} 8} & 
$
\begin{array}{c}
0 \\[2mm]
-1
\end{array} 
$ 
& $\left(
\begin{array}{c}
N_1 \\[2mm]
E_1
\end{array} 
\right)$
$\left(
\begin{array}{c}
N_2 \\[2mm]
E_2
\end{array} 
\right)$
\\ \hline
\end{tabular}
\end{center}
\end{table}
%
Remarkably  just three generations of quark and lepton states survive 
as shown in the table. 
In the bosonic sector, gauge fields of SM in vector states and one Higgs field of SM in the scalar states survive.  
Besides those observed states, new particle:s: {\it one color-singlet neutral vector state {\it S}} and {\it one double-charge color-singlet  spin ${1 \over 2}$ state E} are survived and predicted,  wich can be tested in cosmology and in particle physics experiment.
We will show in the next section that no-go theorem\cite{HLS} for constructing non-trivial $SO(N>8) $SUGRA can be  circumvented by adopting the {\it nonliner (NL)} representation of SUSY, 
i.e. by introducing {\it the degeneracy of space-time} through NLSUSY degrees of freedom. \\[10mm]
\newpage
\noindent
\underline{\bf 2.  Nonlinear-Supersymmetric General Relativity Theory(NLSUSYGR)} \\[3mm]
\indent
For simplicity we discuss $N=1$ without the loss of the generality. \\
The fundamental action 
{\it nonlinear supersymmetric general relativity theory (NLSUSYGR)} 
has been constructed\cite{KS2} by extending the geometric arguments 
of Einstein general relativity (EGR) on Riemann space-time to new space-time inspired by NLSUSY\cite{VA}, 
where tangent space-time is specified  not only by the Minkowski coodinate $x_a$ for $SO(1,3)$ 
but also by the Grassmann coordinate $\psi_\alpha$ for $SL(2,C)$ related to NLSUSY. 
They are coordinates of the coset space  ${super GL(4,R) \over GL(4,R)}$ 
and  can be interpreted as NG fermions associated with the spontaneous breaking of ${\it super}$-$GL(4,R)$ down to $GL(4,R)$. 
The NLSUSYGR action\cite{KS2} is given by 
\begin{eqnarray}
& & L_{NLSUSYGR}(w) =- {c^4 \over 16{\pi}G} \vert w \vert \{\Omega(w) + \Lambda \}, 
\label{SGM}
\\[2mm]
& & \hspace*{7mm} 
\vert w \vert = \det w^a{}_\mu = \det \{e^a{}_\mu + t^a{}_\mu(\psi)\}, 
\nonumber \\[.5mm]
& & \hspace*{7mm} 
t^a{}_\mu(\psi) = {\kappa^2 \over 2i}(\bar\psi \gamma^a \partial_\mu \psi 
- \partial_\mu \bar\psi \gamma^a \psi), 
\label{Lw}
\end{eqnarray}
where $G$ is the Newton gravitational constant, $\Lambda$ is a ({\it small}) cosmological term and 
$\kappa$ is an arbitrary constant of NLSUSY with the dimemsion (mass)${^{-2}}$.   
$w^a{}_\mu(x)$ $= e^a{}_\mu + t^a{}_\mu(\psi)$ and 
$w^{\mu}{_a}$ = $e^{\mu}{_a}
- t{^{\mu}}_a + t{^{\mu}}_{\rho} t{^{\rho}}_a - t{^{\mu}}_{\sigma}t{^{\sigma}}_{\rho} t{^{\rho}}_a  
+ t{^{\mu}}_{\kappa}t{^{\kappa}}_{\sigma}t{^{\sigma}}_{\rho}t{^{\rho}}_a  $ 
which terminate at $O(t^{4})$ for $N=1$ are the invertible {\it unified vierbeins} of new space-time.
{} $e^a{}_\mu$ is the ordinary vierbein of EGR for the local $SO(1,3)$ and    
$t^a{}_\mu(\psi)$ is the mimic vierbein analogue (actually the stress-energy-momentum tensor) of NG fermion $\psi(x){}$ for the local $SL(2,C)$. 
\newpage
\noindent
(We call later $\psi(x){}$  $superon$ as the hypothetical fundamental  spin $1 \over 2$  particle of the supercurrent of the global NLSUSY invariancr and quantized canonically in compatible with the superalgebra\cite{KS4}.)  
$\Omega(w)$ is the the unified Ricci scalar curvature  computed in terms of the {\it unified vierbein}  $w^a{}_\mu(x)$ of new space-time.  
Interestingly Grassmann degrees of freedom induce the imaginary part of 
the unified vierbein $w^a{}_\mu(x)$,  which represents straightforwardly the fermionic matter contribution.
Note that $e^a{}_\mu$ and $t^a{}_\mu(\psi)$ contribute equally to the curvature of spac-time,  which may be regarded as the Mach's principle in ultimate space-time. 
(The second index of mimic vierbein $t$, e.g. $\mu$ of $t^a{}_\mu$,  means the derivative $\partial_{\mu}$.)
$s_{\mu \nu} \equiv w^a{}_\mu \eta_{ab} w^b{}_\nu$ and 
$s^{\mu \nu}(x) \equiv w^\mu{}_a(x)w^{\nu a}(x)$ 
are {\it unified metric tensors} of new spacetime.  \par
$L_{NLSUSYGR}(w)$ (\ref{Lw}) is invariant under the following NLSUSY transformations\cite{ST3}:
\begin{equation}
\delta^{NL} \psi ={1 \over \kappa^{2}} \zeta + 
i \kappa^{2} (\bar{\zeta}{\gamma}^{\rho}\psi) \partial_{\rho}\psi,
\quad
\delta^{NL} {e^{a}}_{\mu} = i \kappa^{2} (\bar{\zeta}{\gamma}^{\rho}\psi)\partial_{[\rho} {e^{a}}_{\mu]},
\label{newsusy}
\end{equation} 
where $\zeta$ is a constant spinor parameter and  $\partial_{[\rho} {e^{a}}_{\mu]} = 
\partial_{\rho}{e^{a}}_{\mu}-\partial_{\mu}{e^{a}}_{\rho}$, 
which close on $GL(4,R)$, i.e. new NLSUSY (\ref{newsusy}) is the square-root of ${GL(4,R)}$; 
\begin{equation}
[\delta_{\zeta_1}, \delta_{\zeta_2}] \psi
= \Xi^{\mu} \partial_{\mu} \psi,
\quad
[\delta_{\zeta_1}, \delta_{\zeta_2}] e{^a}_{\mu}
= \Xi^{\rho} \partial_{\rho} e{^a}_{\mu}
+ e{^a}_{\rho} \partial_{\mu} \Xi^{\rho},
\label{com1/2-e}
\end{equation}
where 
$\Xi^{\mu} = 2i\kappa (\bar{\zeta}_2 \gamma^{\mu} \zeta_1)
      - \xi_1^{\rho} \xi_2^{\sigma} e{_a}^{\mu}
      (\partial_{[\rho} e{^a}_{\sigma]})$.
and induce the following GL(4,R) transformations on the unified vierbein  $w{^a}_{\mu}$ and the metric tensor $s_{\mu\nu} $
\begin{equation}
\delta_{\zeta} {w^{a}}_{\mu} = \xi^{\nu} \partial_{\nu}{w^{a}}_{\mu} + \partial_{\mu} \xi^{\nu} {w^{a}}_{\nu}, 
\quad
\delta_{\zeta} s_{\mu\nu} = \xi^{\kappa} \partial_{\kappa}s_{\mu\nu} +  
\partial_{\mu} \xi^{\kappa} s_{\kappa\nu} 
+ \partial_{\nu} \xi^{\kappa} s_{\mu\kappa}, 
\label{newgl4r}
\end{equation} 
where  $\xi^{\rho}=i \kappa^{2} (\bar{\zeta}{\gamma}^{\rho}\psi)$, \\ 
NLSUSY GR action (\ref{SGM}) possesses promissing large symmetries\cite{ST3,ST4} 
isomorphic to $SO(N)$ ($SO(10)$) SP group; 
namely, $L_{NLSYSYGR}(w)$ is invariant under the spacetime symmetries: 
\begin{center}
$[{\rm new \ NLSUSY}] \otimes [{\rm local \ GL(4,R)}] \otimes [{\rm local \ Lorentz}] $
\end{center}
and 
under the internal symmetries :
\begin{center}
$[{\rm global} SO(N)] \otimes [{\rm local} U(1)^N]$ 
\end{center}
in case of $N$ superons $\psi^{i}, i=1,2,\cdots,N$ 
Note that the no-go theorem is overcome (circumvented) in a sense that 
the nontivial $N$($N > 8)$-extended SUSY with gravity  has been constructed in the NLSUSY invariant way. \\[7mm]
\noindent
{\bf \underline { 3. Big Collapse(BC) of ultimate space-time(NLSUSYGR)}}\\[3mm]
\indent
New {(\it empty)} {} space-time  described  
by NLSUSYGR action $L_{NLSUSYGR}(w)$ 
is unstable due to NLSUSY structure of tangent space-time and 
collapses (called {\it Big Collapse}) spontaneously to 
ordinary Riemann space-time with the cosmological term and fermionic matter {\it superon}  (called {\it superon-graviton model (SGM)}).   \\
$L_{SGM}(e, \psi)$ can be recasted formally as the following famlliar form 
\begin{equation}
L_{NLSUSYGR}(w)= L_{SGM}(e,\psi)=-{c^{4} \over 16{\pi}G}\vert w \vert \{ R(e) +{} \Lambda + \tilde T(e, \psi) \},
\label{SGMR}
\end{equation}
where $R(e)$ is the Ricci scalar curvature of ordinary EH action and 
$\tilde T(e,\psi)$ represents the kinetic term and the gravitational interaction of 
superons. 
Notice that $L_{SGM}(e,\psi)$ in Riemann-flat $e{_a}^{\mu}(x) \rightarrow \delta{_a}^{\mu}$ space-time should reproduce   NLSUSY action of Volkov-Akulov,  i.e.,  the arbitrary constant ${\kappa}$ of NLSUSY is fixed to  
\begin{equation}
{\kappa}^{-2} = {c^4 \over 8{\pi}G}{\Lambda}.
\label{kappa}
\end{equation}
Note that BC induces the rapid expansion of the metric of space-time  due to the Pauli principle for fermion {\it superon};
$ds^{2}=s_{\mu\nu}dx^{\mu}dx^{\nu}=[g_{\mu\nu}(e)+h_{\mu\nu}(e,\psi)]dx^{\mu}dx^{\nu}$  
and simultaneously produces massless eigenstates of $SO(N)  sP$ space-time symmetey by the  all possible gravitational composite of supecharges. 
SGM action, whose gravitational evolution ignites Big Bang of the present observed universe.  
NLSUSY scenario predicts the dimension of space-time is {\it four}, 
for space-time supersymmetry for $SO(1,D-1)$ and $SL(d,C)$ requires 
\begin {equation}
{{D(D-1)} \over 2} = 2(d^{2}-1),
\end{equation}
which holds only for {\it $D=4, d=2$}.
 \\[7mm]
%
%
\noindent
{\bf \underline {4. Evolution of NLSUSYGR/SGM}} \\[3mm]
\indent
NLSUSYGR(SGM) with  $\Lambda>0$  evolves toward the true vacuum. 
The gravirty is the  universal attractive force and  creates all possible gravitational composites of superons, which is the same structure as the all possible products of supercharges and corresponds to  (massless) helicity-eigenstates of SO(10)  {\it linear(L)}SUSY sP algebra of asymptotic 
space-time symmetry.
 (Note that the leading term of supercharge is the superon field.)  
This means that all component fields of LSUSY supermultiplet are expressed 
as such composites of superons (called {\it SUSY compositeness}) as familiarLSUSY transformation of LSUSY supermultiplet are reproduced on composite supermultiplet under the NLSUSY transformations of the constituent superons. 
Simultaneously the equivalence of NLSUSY action and the LSUSY action holds (called {\it NL/L SUSY relation}) in the sence that LSUSY action reduces to NLSUSY action  when SUSY compositeness is inserted in LSUSY component fields.  
To see the low energy (vacuum) behavior of $N = 2$ SGM (NLSUSYGR) 
we consider SGM in asymptotic  Riemann-flat space-time, where $N = 2$ SGM reduces to essentially $N = 2$ NLSUSY action. 
We will show the equivalence  (called {\it NL/L SUSY relation}) of $N = 2$ NLSUSY action to  $N = 2$ LSUSY QED action\cite{STT1,STT2}, i.e., 
%
\begin{equation}
{L_{{\rm NLSUSYGR}}({w^{a}}_{\mu})} ={L_{{\rm SGM}}}({e^{a}}_{\mu}. \psi) \rightarrow
{L_{{\rm NLSUSY}}(\psi)} + [{\rm suface\ terms}]
= f_{\xi^{i}} {L_{{\rm LSUSY}}(v^{a},\ D. \cdots)}
\end{equation}
where $f_{\xi^{i}}$ is the function of  vacuum values $\xi^{i}$ of auxiliary fields 
of LSUSY supermultiplet. 
The  relation between LSUSY and NLSUSY is studied in detail in \cite{IK}.
Applying the arguments to $N=2$ NLSUSY  and LSUSY QED theory 
we obtain the  SUSY compositeness for
the LSUSY QED theory\cite{ST6a}:  
For example, the SUSY compositeness for the {\it minimal gauge} vector supermultiplet $(v^a, \lambda^i, A, \phi, D)$ are 
%
%
\begin{eqnarray}
v^a = - {i \over 2} \xi \kappa \epsilon^{ij} 
\bar\psi^i \gamma^a \psi^j \vert w \vert, \ \
\lambda^i = \xi \left[ \psi^i \vert w \vert 
- {i \over 2} \kappa^2 \partial_a 
\{ \gamma^a \psi^i \bar\psi^j \psi^j 
(1 - i \kappa^2 \bar\psi^k \!\!\not\!\partial \psi^k) \} \right], 
\nonumber \\[.5mm] 
A = {1 \over 2} \xi \kappa \bar\psi^i \psi^i \vert w \vert, 
\phi = - {1 \over 2} \xi \kappa \epsilon^{ij} \bar\psi^i \gamma_5 \psi^j 
\vert w \vert, 
%
D = {\xi \over \kappa} \vert w \vert 
- {1 \over 8} \xi \kappa^3 
\partial_a \partial^a ( \bar\psi^i \psi^i \bar\psi^j \psi^j ).  \  \cdots
\label{SSUSYinv1}
\end{eqnarray}

and similar results for the scalar supermultiplet $(\chi. B^{i}, F^{i}, \nu^{i})$ 
of LSUSY QED theory.  
NL/L SUSY relation(equivalence)  (9) is shown explicitly by substituting 
the following SUSY compositeness into the LSUSY QED theory. 
NL/L SUSY relation shows the equivalence(relation) of two theories  
irrespective of the renormalizability.  \par
NL/L SUSY relation $f_{\xi^{i}}=1$ for the {\it non-minimal most  general} gauge vector and scalar supermultiplet  predicts the magnitude of the bare gauge coupling constant  as follows\cite{gaugecp} :   
\begin{equation}
f(\xi,\xi^{i}, \xi_{c})=\xi^2 - (\xi^i)^2e^{-4e\xi_{c}} = 1, \ \ \ i.e.,  \ \ \
{e} = {{\ln({\xi^i{}^2 \over {\xi^2 - 1}}) \over 4 \xi_c}}, 
\label{f-xi}
\end{equation}
where  $\xi$  and $\xi^{i}$ are the magnitudes of the vacuum values of $D$ and $F$, respectively.     

{{Broken LSUSY(QED) gauge theory  is encoded   
in the vacuum of NLSUSY theory   
as composites of NG fermion.} }
SM and GUT may be interpreted as an low energy (effective) gravitational  composite theory of SGM.　\\[10mm]
{\bf \underline {5. Low energy particle physics of NLSUSYGR/SGM}}  \\[3mm]
\indent
{NL/L SUSY relation(equivalence)} gives:
\begin{center}
${L_{N=2{\rm SGM}}  {\longrightarrow}  
L_{N=2{\rm NLSUSY}} + [{\rm suface\ terms}]
= L_{{N=2{\rm SUSYQED}}}}$
\end{center}
in Riemann-flat space-time. 
%
%
The vacuum is given by the minimum of the potential {{$V(A, \phi, B^i, D)$ of 
$L_{N=2{\rm LSUSYQED}}$}}, 
\begin{equation}
V(A, \phi, B^i, D)  =  - {1 \over 2} D^2 + \left\{ {\xi \over \kappa} 
- f(A^2 - \phi^2) + {1 \over 2} e |B^i|^2 \right\} D 
+{e^{2} \over 2}(A^2 + \phi^2)|B^{i}|^{2}.   
\end{equation}
We find the vacuum which has the following  mass spectra for the particle configuration around the true vacuum\cite{ST7,STL}:
\begin{equation}
m_{\hat A}^2 = m_{\lambda^i}^2 = 4 f^2 k^2 = {{4 \xi f} \over \kappa}, \ \
m_{\hat B^1}^2 = m_{\hat B^2}^2 = m_\chi^2 = m_\nu^2 
= e^2 k^2 = {{\xi e^2} \over {\kappa f}}, \ \
m_{v_{a}} = m_{\hat \phi} = 0,
\end{equation}
which produces {\it mass hierarchy} by the factor 
{ ${e \over f}$},
The vacuum breaks SUSY softly  $<A> \neq 0$ and  describes qualitatively  
{lepton-Higgs-U(1) sector analogue of the SM: \
one massive charged Dirac fermion ({$\psi_D{}^c \sim \chi + i \nu$}), \ 
one massive neutral Dirac fermion ({$\lambda_D{}^0 \sim \lambda^1 - i \lambda^2$}), \ 
one massless vector (a photon) ({ $v_a$}), \ 
one charged scalar ({ $\hat B^1 + i \hat B^2$}), \ 
one neutral complex scalar ({$\hat A+ i \hat \phi $}),  \ 
which are {composites of superons}. \par
Revisiting SM and GUT by adopting the SQM (superon-quintet composite model) picture may give new insight into the unsolved problems of the SM. 
For example, adopting the following simple assignment of observed particles:   
$ (e, \nu_{e})$: $\delta^{ab}Q_{a}{Q^{*}}_{b}Q_{m}$, \ 
$(\mu, \nu_{\mu})$: $\delta^{ab}Q_{a}{Q^{*}}_{b}\epsilon^{lm}Q_{l}Q_{m}{Q^{*}_{n}}$, \
$(\tau, \nu_{\tau})$:$\epsilon^{abc}Q_{b}Q_{c}\epsilon^{ade}$${Q^{*}}_{d}{Q^{*}_{e}Q_{m}}$,  \
$(u,d)$: $\epsilon^{abc}Q_{b}Q_{c}Q_{m}$,  \
$(c,s)$: $\epsilon^{lm}Q_{l}Q_{m}\epsilon^{abc}Q_{b}Q_{c}Q^{*}{_n}$, \
$(t,b)$: $\epsilon^{abc} Q_{a}Q_{b}Q_{c}{Q^{*}}_{d}Q_{m}$,  \
$Gauge \ Boson$: $Q_{a}{Q^{*}}_{b}, \cdots$,        \
$Higgs  \ Boson$: $\delta^{ab}Q_{a}{Q^{*}}_{b}Q_{l}{Q^{*}}_{m}$, $\cdots$ \.    
most of the Feynman diagrams of SM and GUT are reproduced by composite multiple superon-line SQM diagrams for SM particles, e.g., the SQM diagram for $\beta$-decay  is depicted below.   \\[10mm]
\setlength{\unitlength}{1mm}
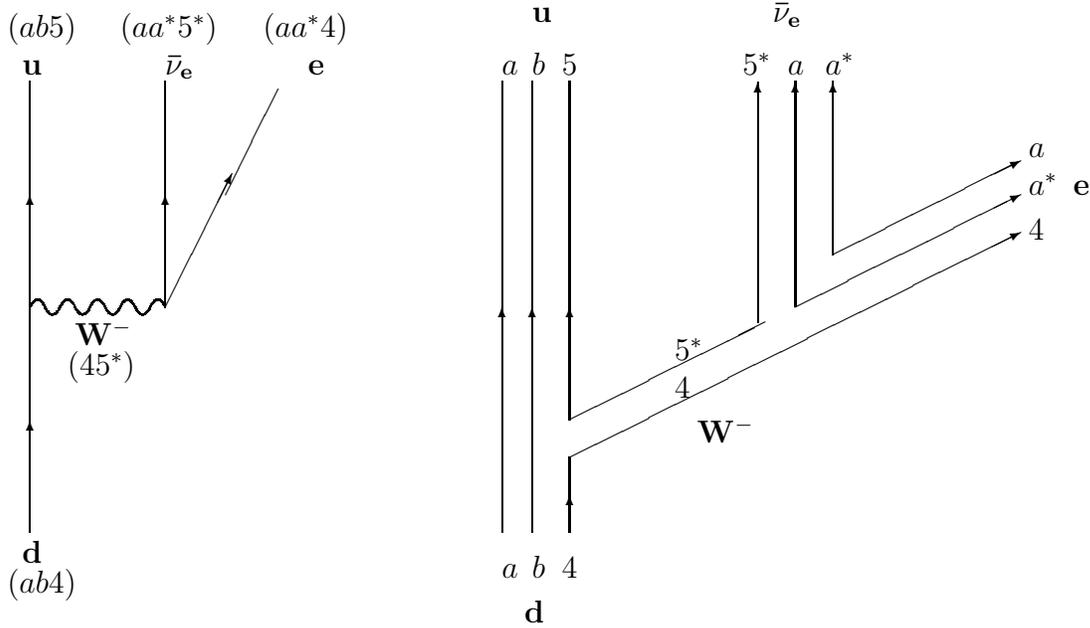
\begin{picture}(70,70)

\put(3,-4){$\bf d$}
\put(1,-8){$(ab4)$}
\put(1,66){$(ab5)$}
\put(3,61){${\bf u}$}
\put(4,0){\vector(0,1){15}}
\put(4,15){\vector(0,1){30}}
\put(4,45){\line(0,1){15}}

\put(22,30){\vector(0,1){15}}
\put(22,45){\line(0,1){15}}
\put(16,66){$(aa^{*}5^{*})$}
\put(22,61){${\bf {\bar{\nu}_{e}}}$}
\put(22,30){\vector(1,2){8.9}}
\put(30,45){\line(1,2){7}}
\put(41,61){$\bf e$}
\put(35,66){$(aa^{*}4)$}
\put(10,25){$\bf {W^{-}}$}
\put(9,21){$(45^{*})$}

\qbezier(4,30)(5,32)(6,30)

\qbezier(6,30)(7,28)(8,30)

\qbezier(8,30)(9,32)(10,30)

\qbezier(10,30)(11,28)(12,30)

\qbezier(12,30)(13,32)(14,30)

\qbezier(14,30)(15,28)(16,30)

\qbezier(16,30)(17,32)(18,30)

\qbezier(18,30)(19,28)(20,30)

\qbezier(20,30)(21,32)(22,30)

\end{picture}
%
%
%
%
%
%
\setlength{\unitlength}{1mm}
\begin{picture}(60,60)
\put(-6,0){\vector(0,1){30}}
\put(-6,30){\line(0,1){30}}
\put(-6,-6){$a$}
\put(-6,61){$a$}
\put(-2,0){\vector(0,1){30}}
\put(-2,30){\line(0,1){30}}
\put(-2,-6){$b$}
\put(-2,61){$b$}
\put(-2,68){\bf u}
\put(2,-6){$4$}
\put(-3,-12){\bf d}
\put(3,0){\vector(0,1){5}}
\put(3,5){\line(0,1){5}}
\put(3,10){\vector(2,1){60}}

\put(2,61){$5$}
\put(3,15){\vector(0,1){15}}
\put(3.,30){\line(0,1){30}}
\put(3,15){\line(2,1){26}}

\put(28,28){\vector(0,1){32}}

\put(38,37){\vector(2,1){25}}

\put(33,30){\vector(0,1){30}}

\put(33,30){\vector(2,1){30}}

\put(38,37){\vector(0,1){23}}

\put(17,18){$4$}
\put(17,23){$5^{*}$}
\put(26,61){$5^{*}$}
\put(32,61){$a$}
\put(37,61){$a^{*}$}
\put(70,45){\bf e}
\put(30,68){${\bf {\bar{\nu}_{e}}}$}   
\put(20,12){$\bf {W^{-}}$}
\put(64,50){$a$}
\put(64,45){$a^{*}$}
\put(64,39){$4$}
\end{picture}  \\[15mm]

However, interestingly the (leading order) diagrams of the proton decay, FCN current, $\cdots$, etc. are forbidden(stable proton?) by the superon selection rule at the vertex. 
SQM  gives new viewpoints of CP violation, dark matter, dark energy and predicts two new color-singlet heavy  particles: one neutral vector boson $S$: $\delta^{ab}Q_{a}{Q^{*}}_{b}$ and one double-charge spin ${1 \over 2}$ fermion $E^{\pm2}$:  $\epsilon^{abc}Q_{a}Q_{b}Q_{c}\epsilon^{lm}{Q^{*}}_{l}{Q^{*}}_{m}$ 
whose evidences may be observed in the high energy/cosmic ray  experiments..     
 \par
NL/L SUSY reation for  $N>3$ ( $N=3$ SYM is passed.) is yet to be done.
The potential of NLSUSYGR/SGM scenario has been discussed briefly in the toy model, though many open questions are left.   \\[5mm]
\noindent
{\bf \underline {6. Cosmological implications of NLSUSYGR/SGM}} \\[3mm]

BC of ultimate space-time described by NLSUSYGR to SGM on  ordinary Riemann space-time may give new insight into the evolution of space-time and matter around BB\cite{TOL}. \\
\indent
The variation of SGM action $L_{\rm SGM}(e,\psi)$ with respect to  ${e^{a}}_{\mu}$ yields  \\
{ Einstein equation equipping with  matter and cosmological term}: 
\begin{equation}
{ R_{\mu\nu}(e)-{1 \over 2}g_{\mu\nu}R(e)=
{8{\pi}G \over c^{4}} \{ \tilde T_{\mu\nu}(e,{\psi})-{g_{\mu\nu}{c^{4}\Lambda \over 8{\pi}G}} \}}.
\label{SGMEQ}
\end{equation}
where $\tilde T_{\mu\nu}(e,{\psi})$ abbreviates the stress-energy-momentum of superon(NG fermion) 
including the gravitational interaction. \
Note that { the cosmological term} ${ -{c^{4}\Lambda \over 8{\pi}G}}$ can be interpreted as 
{\it  the negative energy density of space-time},  \
i.e. {the dark energy density ${{\rho}_{D}}$}. \\
Big collapse(BC) $L_{\rm NLSUSYGR}({w^a}_{\mu}) \rightarrow  L_{\rm SGM}(e,\psi)$  may induce  3 dimensional  ${inflational}$  expansion} of space-time 
 { by Pauli principle}: 
\begin{center}
$ds^{2}=s_{\mu\nu}(x)dx^{\mu}dx^{\nu}=\{ g_{\mu\nu}(e)+{h_{\mu\nu}(e,\psi)} \} dx^{\mu}dx^{\nu}$. \\[2mm]
$\{\psi,\bar {\psi} \}=0  \  \Rightarrow  \  \{\psi(x), \bar {\psi} (y) \} = \delta^{(3)}(\bf{x-y})$,
\end{center}
where ${h_{\mu\nu}(e,\psi)}$ represents the contribution of fermion matter.
\noindent
BC produces  composite (massless) eigenstates  of  SO(N) sP algebra  
{ due to the universal attractive force of graviton}, 
which is {the ignition of the Big Bang(BB) SM scenario}.\ \
As shown in the toy model, the vacuum of { the composite SGM scenario} may explain naturally { observed mysterious (numerical) relations: } 
\begin{center}
$dark \ energy\ density \ \rho_{D} \sim O(\kappa^{-2})  \sim {m_\nu}^4 \sim (10^{-12}GeV)^{4}$
\end{center}
provided { $\lambda_D{}^0$ is identified with neutrino and $f\xi \sim O(1)$}.   
    \\[5mm]
\noindent
{\bf \underline{7. Summary}}  \\[2mm]
The hindsight of the discussions on NLSUSYGR scenario suggests a new possibility of SUSY unification of space-time and matter, where NLSUSY(NG fermion) and GR(graviton) play  essential roles before /around BB.of the universe. However there remains many open questions, e.g., the direct linearization of $L_{NLSUSYGR}(e,\psi)$, the significance of spin 3/2 version 
$L_{NLSUSYGR}(e,{\psi}_{\mu})$\cite{ST8}, etc.  \\[3mm]

The author would like to thank Y. Tanii,  J. Sato and otherr colleagues of the theory group of particle physics of Saitama University for their warm hospitality.  \\[5mm]
%


%

\end{document}